\documentclass{egpubl}
\usepackage{eg2026}
 
\ShortPresentation      
\usepackage[T1]{fontenc}
\usepackage{dfadobe}

\usepackage{cite}  
\BibtexOrBiblatex
\electronicVersion
\PrintedOrElectronic
\ifpdf \usepackage[pdftex]{graphicx} \pdfcompresslevel=9
\else \usepackage[dvips]{graphicx} \fi

\usepackage{egweblnk} 

\usepackage{array}
\usepackage{graphicx}
\usepackage{subfig}
\usepackage{xcolor}
\usepackage{xspace}

\usepackage{amssymb}

\definecolor{todocolor}{rgb}{0.9,0.1,0.1}

\newcommand{\algformat}[1]{\textsc{#1}\xspace}
\newcommand{\poh}{\algformat{Paris Opera House}}
\newcommand{\sample}{\algformat{O3DE Multiplayer Sample}}
\newcommand{\hdc}{\algformat{Exhibition}}
\newcommand{\singlegpu}{\algformat{SingleGPU}}
\newcommand{\dualgpu}{\algformat{DualGPU}}
\newcommand{\quadgpu}{\algformat{QuadGPU}}
\newcommand{\fortis}{\algformat{Fortis}}

\title{Capsule: Efficient Player Isolation for Datacenters}

\author[Z. Du et al.]
{\parbox{\textwidth}{\centering Z. Du\orcid{0009-0002-0100-7565},
        N. Davari\orcid{0009-0008-4288-1922},
        L. Li\thanks{Work done while at Huawei}\orcid{0009-0002-3514-8673},
        W. S. Loi\orcid{0000-0003-3995-2052},
        N. Kodirov\orcid{0000-0002-9674-3559}
        }
        \\
{\parbox{\textwidth}{\centering Huawei Technologies, Vancouver, BC, Canada}
}
}

\usepackage[labelfont=bf, textfont=it]{caption}

\begin{document}

\teaser{
 \includegraphics[width=0.99\linewidth]{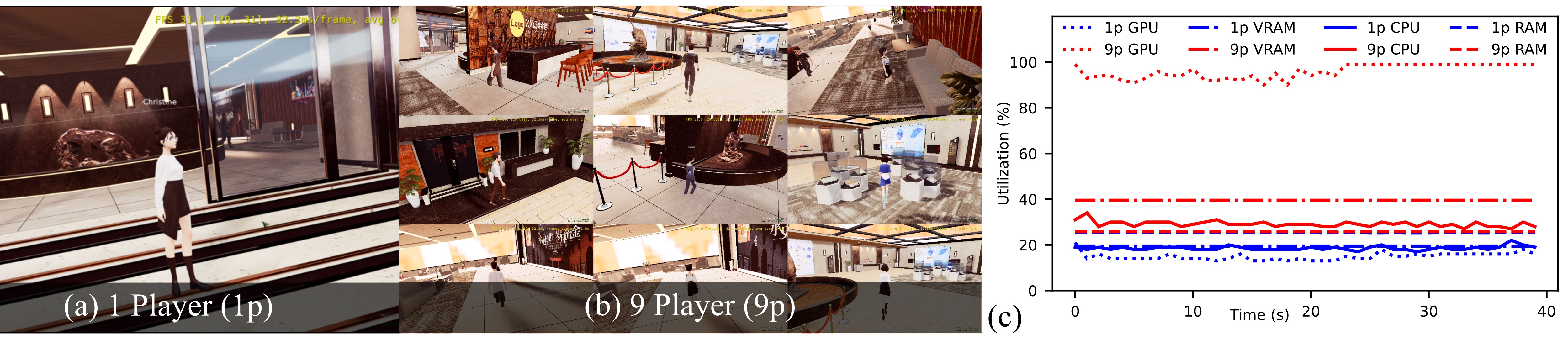}
 \centering
 \caption{An example application for a museum digital twin:
   (a) no Capsule, one player per engine,
   (b) with Capsule, nine players per engine,
   (c) compares datacenter resource utilization with and without Capsule.
   Capsule provides lightweight and efficient player isolation.}
\label{fig:teaser}
}

\maketitle

\begin{abstract}

We introduce Capsule, a mechanism for seamlessly sharing
datacenter resources across multiple players.
It decouples player-local and global states to
achieve isolation and to maximize cross-player sharing.
Our evaluations show that Capsule increases datacenter resource utilization
by accommodating up to 2.25x more players
without degrading the user experience.
This improvement stems from Capsule consuming up to
1.43x less GPU, 3.11x less VRAM,
3.7x less CPU, and 3.87x less RAM
compared to the baseline.
We evaluated Capsule across four applications and various hardware configurations,
including three distinct servers and a multi-server cluster. These results
demonstrate that the Capsule design is portable to other game engines.


\end{abstract}
\section{Introduction}

Cloud gaming is attractive for both players and cloud providers.
For players, it alleviates deployment costs.
They no longer need to own the latest hardware (e.g., GPU)
to play high-quality games.
For providers, the goal is to generate revenue while delivering the highest gaming quality.
The more players, the higher the revenue.

However, scaling cloud gaming capacity within datacenters remains challenging.
Part of this challenge arises from
games having diverse \textit{shapes} and \textit{sizes}.
Shapes correspond to the diverse resources that games consume,
such as CUDA cores, RT cores, and Tensor cores in GPUs,
in addition to the host CPU and RAM.
Sizes correspond to the differing amounts of these resources that games consume,
e.g., a graphics-intensive game consumes the entire GPU,
while a graphics-light game consumes only a fraction of that GPU.
Another challenge arises from the diverse hardware found in datacenters.
Datacenters often house servers with several generations of CPUs and GPUs.
For example, a server with an older GPU
might accommodate only one player,
while a server with the latest GPU accommodates dozens.
Thus, cloud providers need a mechanism to
share GPUs, as well as other resources, across multiple players.

We propose player-level multiplexing.
A player in a graphics-heavy application will continue consuming the entire GPU.
However, when a GPU has sufficient capacity to accommodate two or more players
in a multiplayer game,
its resources will be multiplexed across these players.
We designed, implemented, and evaluated
Capsule: an in-game-engine player isolation mechanism
for multiplayer games.
Capsule allows cross-player \textit{sharing}.
For example, when two players enter a room and have a shared game asset
in their view, we can reuse the asset geometry across these two players
without players noticing.
Sharing allows amortization, i.e., sublinear growth in datacenter utilization
for a linear increase in player count: a phenomenon we call a
\textit{sublinear resource footprint}.

We implemented Capsule in O3DE: our production game engine.
It satisfies four requirements for player isolation in the cloud:
\begin{itemize}
\item {\textbf{R1: Transparent}}: Players should be unaware of other players
sharing cloud resources. A player's experience---such as input latency,
output streaming quality, and frames-per-second (FPS)---should
not degrade due to other players.
\item {\textbf{R2: Compatible}}: Player isolation should not necessitate
significant modifications to existing applications;
ideally, it should require no changes at all.
Furthermore, the development workflow for new applications should not change.
\item {\textbf{R3: Lightweight}}: The isolation mechanism itself should not
consume significant system resources, such as CPU and RAM.
\item {\textbf{R4: Efficient}}: Maximize cross-player sharing.
For example, the resource footprint (e.g., CPU) of the second player should be
less than that of the first player, because
the second player can reuse parts of
the computation results from the first player.
\end{itemize}

In summary, we make the following contributions:
\textbf{(1)} We define four practical requirements (R1--R4) for player isolation
that are broadly applicable across cloud providers.
\textbf{(2)} We propose Capsule, a novel in-engine mechanism that implements
these requirements by decoupling player-local and global states to maximize
cross-player resource sharing.
\textbf{(3)} We provide a comprehensive evaluation of Capsule using
applications with diverse performance characteristics and
various hardware specifications.
\section{Related Work}

\begin{table}[t]
  \centering
  \caption{Existing work: academic papers and real deployments.}
  \label{tab:lit}
  \begin{tabular}{|p{3.4cm}|c|c|c|}
    \hline
    \textbf{} & \textbf{Console} & \textbf{O3DE} & \textbf{Capsule} \\
    \hline
    Remote & x & \checkmark & \checkmark  \\
    \hline
    Compute \& Mem. Sharing & \checkmark & x & \checkmark\\
    \hline
  \end{tabular}
\end{table}

\autoref{tab:lit} broadly categorizes existing work
into three groups across two dimensions.
The \textit{Console} group represents deployments with user hardware,
such as the PS5, Xbox, and PC. Here, frames are generated on the user hardware,
i.e., no heavy computation occurs in the cloud.
The user console's compute and memory resources are shared across
players (e.g.,~\cite{it-takes-two}).
On the other hand, in the second group, labeled \textit{O3DE},
frames are generated in the cloud,
but cloud resources cannot be shared
across players.
This is the case for any game engine running in the cloud
(e.g., Unreal Engine and Unity),
although we label the group \textit{O3DE}.
Capsule (i.e., O3DE with Capsule)
enables cross-player resource sharing in the cloud.

A large body of literature, including CloudLight~\cite{CloudLight}
and Weinrauch et al.~\cite{osc}, falls within the Capsule category.
These systems generate full or partial frames in the cloud and
share the resources required to generate
those frames across multiple players.
However, these works primarily demonstrate the feasibility of
offloading rendering computations to the cloud without
addressing the practical requirements (R1--R4).
\section{Design and Implementation}

\begin{figure}[t]
  \centering
  \includegraphics[width=0.99\linewidth]{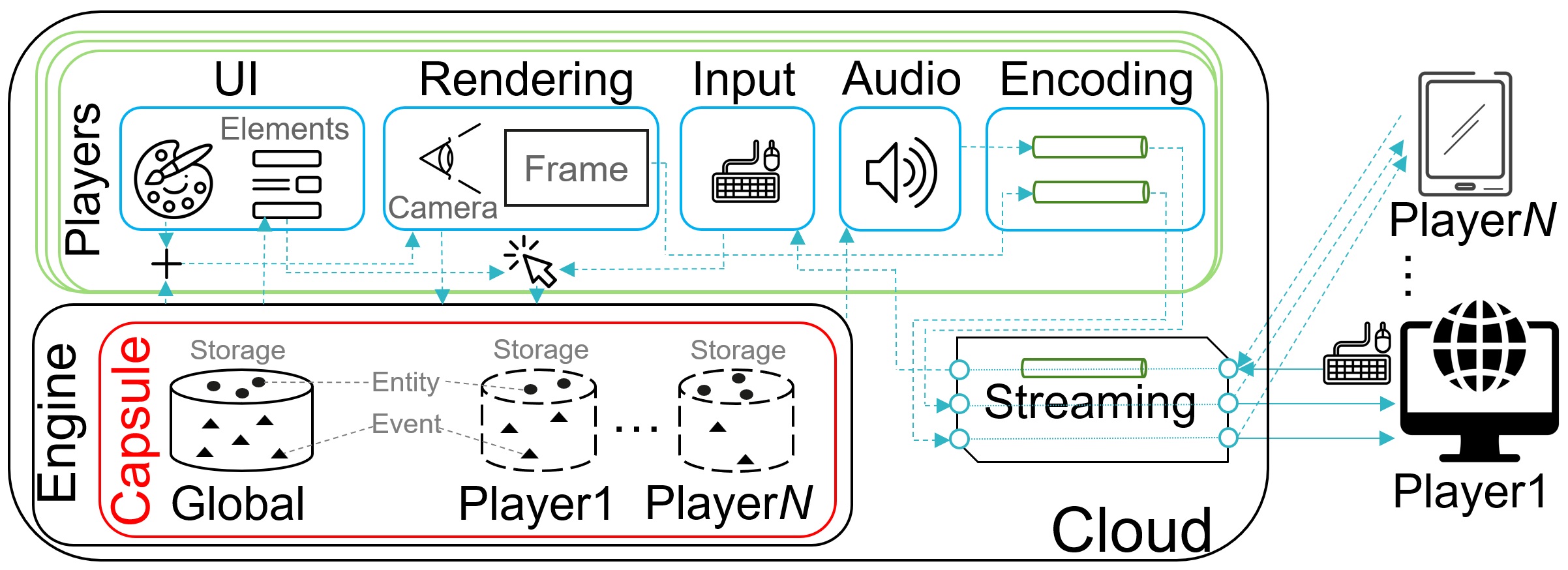}
  \caption{Capsule-based cloud architecture.
  Capsule Storage manages player state,
  which is a key part of player isolation.}
  \label{fig:capsule}
\end{figure}

\autoref{fig:capsule} shows the Capsule architecture,
along with other essential modules in a cloud deployment.
Capsule is a new module in O3DE that
communicates with different system components, such as
audio system, input system, rendering system,
and game logic (event system).
Capsule leverages the Entity--Component--System (ECS) architecture in O3DE.
ECS makes it convenient to represent game world objects.
An ECS-based engine contains \textit{entities} that have data \textit{components}
and \textit{systems} to operate on those components.

As shown in~\autoref{fig:capsule}, players connect
to the cloud over a wide area network (e.g., the Internet).
The Streaming module serves as the entry point, creating a distinct
game session for each player to isolate their respective inputs
(e.g., keyboard) and outputs (e.g., video streams).
It routes player-specific input to an isolated game session containing
dedicated UI, Rendering, Input, Audio, and Encoding states.
While these states are managed within the engine, the application
perceives them as fully isolated, as if each player were running
a dedicated engine.
In reality, Capsule-based O3DE enables all players to share a
single engine.

Capsule isolates player-specific inputs, outputs, and behaviors through 
Capsule Storage. Capsule Storage manages two constructs: entities and events.
Entities are game objects, such as a car, a light, or an avatar.
Events, or gameplay events, include in-game events
such as running, jumping, or exploding.
They are predesigned by the game developer and are written in the game script.
Entities and events determine
the behavior of each player and of the global environment.
They also determine the final rendered frame.

Capsule has two types of Storage: global and local.
There is only one global storage in the entire engine.
All players share the entities and events inside the global storage.
There are one or more local Capsule Storages: one for each player.
Entities inside the local storage are visible only to the storage-owner player.
Upon the API call for the first player, predefined global entities 
are initialized unless they have already been loaded via a cache.
When a player joins, all entities specified by the developer for
the initial level are spawned.
For subsequent players, predefined global entities are not
re-instantiated; instead, references to the original entities are
created within each player's local storage.
These references ensure that the local gameplay logic remains
consistent with the standard game-level design of a non-Capsule engine.
Capsule isolates player-specific tasks at runtime by
directing player-specific entities and events to
the respective local storage.
\autoref{appx:implementation} elaborates on
player entity tracking using global and local storage.

We ported four applications to the Capsule-based O3DE
to validate our design.
All four applications fit within our entity and event tracking systems.
\autoref{fig:teaser} shows only one of them for brevity.
\autoref{fig:teaser-extended} shows the other two.
Our experience with these four applications confirms Capsule's full compatibility (\textbf{R2}).
\section{Evaluation}
\label{sec:eval}

\begin{table}[t]
  \centering
  \caption{Servers used for Capsule evaluation.
  All servers use NVIDIA GPUs.
  \textit{(*Exact GPU model is not disclosed.
  VRAM ``24+'' means it has more VRAM than the above two.
  \fortis is our custom label, meaning \textbf{strong} in Latin.)}}
  \label{tab:workstations}
  \begin{tabular}{|p{1.3cm}|p{1.7cm}|p{1.38cm}|p{1.85cm}|}
    \hline
    \textbf{} & \textbf{\singlegpu} & \textbf{\dualgpu} & \textbf{\quadgpu} \\
    \hline
    CPU & \raggedright AMD Ryzen 7 5800X & \raggedright Intel Core i7-13700K & Threadripper PRO 5975WX \\
    \hline
    \# of Cores & \multicolumn{1}{c|}{8} & \multicolumn{1}{c|}{16} & \multicolumn{1}{c|}{32} \\ 
    \hline
    RAM & \multicolumn{1}{c|}{32 GB} & \multicolumn{1}{c|}{64 GB} & \multicolumn{1}{c|}{256 GB} \\ 
    \hline
    GPU & \raggedright GeForce RTX 4090 & \raggedright GeForce RTX 3090 & \fortis* \\
    \hline
    VRAM & \multicolumn{1}{c|}{24 GB} & \multicolumn{1}{c|}{24 GB} & \multicolumn{1}{c|}{24+} \\ 
    \hline
    \# of GPUs & \multicolumn{1}{c|}{1} & \multicolumn{1}{c|}{2} & \multicolumn{1}{c|}{4} \\ 
    \hline
  \end{tabular}
\end{table}

We evaluated Capsule on diverse datacenter hardware.
We used three different servers:
one with a single GPU, one with two GPUs, and one with four GPUs, as described in
\autoref{tab:workstations}. All servers run the Windows OS to
faithfully reproduce our production environment.
We fixed the application FPS to 30,
a common minimum threshold.
We read the system-wide utilization levels of the
GPU, VRAM, CPU, and RAM resources every second.
The value for each second is the average utilization
during that one-second interval, which is
consistent with Windows performance counters.

We compare the cloud server resource consumption of Capsule against the baseline.
For the baseline, we implemented process-level isolation
(i.e., a separate game engine process for each player).
In the baseline, we launched the game server
and then launched game clients one by one,
measuring the server utilization as players were added.
The client-side evaluation is identical between Capsule and the baseline,
but on the server side, after the first player,
we kept adding players to the same client process
(running in the cloud server),
rather than creating a separate process per player (in the cloud server).
This is consistent with the Capsule design in~\autoref{fig:capsule}.
(We were unable to use production-level alternative virtualization techniques,
such as the NVIDIA RTX Virtual Workstation (vWS),
as the baseline due to licensing restrictions.)

We evaluated Capsule's applicability to diverse datacenter workloads
by running three different
applications---\poh, \sample~\cite{o3de-game}, and \hdc---on the \singlegpu server.
For brevity,
\autoref{fig:teaser} shows results for only the \hdc application
on the \singlegpu server, which has the strongest GPU.
We monitored resource utilization for 40 seconds of gameplay time.
The supplementary material provides gameplay footage
and \autoref{appx:method} details our evaluation methodology.

\autoref{fig:teaser}(a) and \autoref{fig:teaser}(b) show
the game server view of the \hdc application
with a single player (1p) and nine players (9p), respectively.
\autoref{fig:teaser}(c) shows resource utilization for these two environments.
With a single player, GPU, VRAM, and CPU utilization are each $\approx$20\%,
while RAM utilization is $\approx$24\%.
Capsule takes advantage of the remaining capacity
to allocate 8 more players.
The GPU becomes a bottleneck with the 10th player,
dropping the player FPS below the threshold (30).
At this point, we stop allocating more players.
As \autoref{fig:teaser}(c) shows, GPU utilization increases
from $\approx$20\% with 1 player to $\approx$99\% with 9 players.

\begin{figure}
  \centering
  \includegraphics[width=0.99\linewidth]{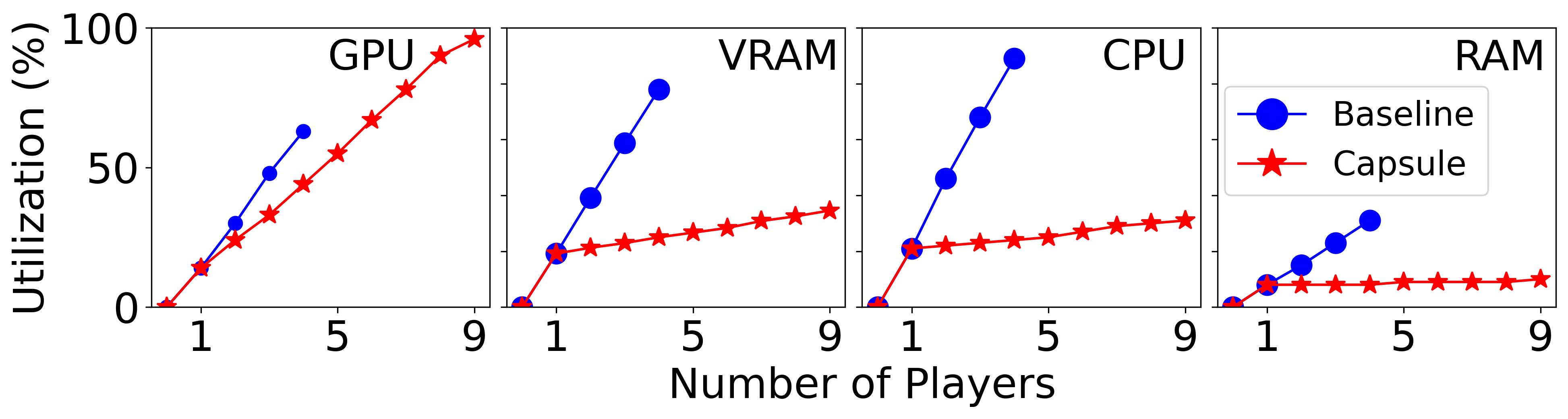}
  \caption{Capsule's scalability with more players.
  Capsule and the baseline host up to 9 and 4 players, respectively, on the \hdc application.
  Capsule accommodates up to 2.25x more players thanks to the sublinear resource increase
  per added player.}
  \label{fig:scale}
\end{figure}

\autoref{fig:scale} shows the average resource utilization increase
as players join the \hdc application in~\autoref{fig:teaser}(c).
\autoref{fig:scale} shows that overall GPU utilization
increases by $\approx$10 percentage points per additional player, after the first one.
VRAM, CPU, and RAM show a similar trend.
Unlike in~\autoref{fig:teaser}, in~\autoref{fig:scale}
we kept adding players in the baseline
(i.e., adding another game-engine process for each player).
The baseline sharply dropped to single-digit FPS
after 4 players due to CPU contention.
At the peak of the baseline (4 players),
Capsule used
\textbf{1.43x less GPU}, \textbf{3.11x less VRAM},
\textbf{3.7x less CPU}, and \textbf{3.87x less RAM}
than the baseline.
These resource savings allow Capsule to accommodate more players.
While the baseline's resource consumption
scales linearly with the number of players,
Capsule requires a sublinear amount of additional VRAM, CPU, and RAM for each
player beyond the first. Capsule's GPU utilization
could be further optimized by incorporating more shareable rendering
techniques, such as shadowmaps, global illumination, and cross-view
diffuse and effect sharing~\cite{osc}.
In some games, cross-player state sharing is insignificant
but game asset sharing is significant.
In these cases, Capsule transparently (to developers)
provides high VRAM and RAM amortization
despite minimal CPU amortization gains.

\autoref{fig:teaser}(c) contrasts the utilization levels
for the baseline (1p) and Capsule (9p), which we call \textit{delta}.
The delta is the highest for the GPU resource.
This delta is smaller but is more \textbf{significant}
in other resources (VRAM, CPU, RAM).
If the delta was the smallest (i.e., is zero), the 1p and 9p lines would overlap.
This would mean that the 9p utilization of that resource is identical to that of 1p;
i.e., the additional 8 players came for free (for that resource).
In~\autoref{fig:teaser}(c), this is almost the case for VRAM,
which is also reflected in~\autoref{fig:scale}.
\autoref{fig:scale} shows that RAM consumption stays relatively flat
as more players are added.
It also demonstrates that the slope of the scaling curve varies
by resource, with the GPU exhibiting the highest degree of
sublinearity and RAM the lowest.
This is because Capsule achieves the greatest degree of
cross-player RAM sharing within the \hdc application.

We evaluated Capsule with two other applications
and diverse hardware. We summarize the results.
One of the two other applications is from our production,
called \poh: a digital twin of the Paris Opera House.
The second application is an open-source multiplayer
shooting game: \sample~\cite{o3de-game}.
Both applications are more graphics-intensive than \hdc.
In \poh, Capsule accommodates
only two players (before hitting the FPS threshold)
and only four players in \sample.
Overall, Capsule's sublinearity benefits also hold
in these applications,
e.g., $\approx$10 percentage points lower GPU utilization per player with Capsule in
the most graphics-intensive \poh application.

In our hardware diversity experiments,
we evaluated \hdc on the three servers in~\autoref{tab:workstations}
as well as on a two-server cluster that connects
\dualgpu and \quadgpu servers over the network.
\autoref{fig:teaser} and \autoref{fig:scale} already show results
for the \singlegpu. The sublinearity trends also hold for the
\dualgpu (hosts up to 8 players)
and \quadgpu (hosts up to 16 players) servers,
where the GPU resource has the highest delta,
followed by the CPU and then VRAM,
with the RAM resource having the lowest delta.
These results also hold for the two-server cluster,
which hosts up to 24 players in aggregate.
\autoref{appx:eval} further elaborates on
application variety and hardware diversity experiments.
\section{Limitations and Discussion}

Capsule has three major limitations:
(1) a CPU bottleneck,
(2) performance isolation, and (3) fate-sharing.
The current implementation of Capsule
focuses on multiplexing GPU resources
because GPUs are expensive and are the main source of bottlenecks
in our production.
However, for some applications, the bottleneck shifts
to the CPU or other resources.
In \poh, for example, Capsule accommodates only two players
because the third player drops the FPS below the threshold (30)
due to a CPU bottleneck.
Capsule currently has one main thread for all players,
which runs on a single CPU core.
That thread becomes the bottleneck.
We can extend Capsule to use different CPU cores
for different players.

Capsule currently offers functional isolation;
e.g., one player jumping does not interfere with
another player jumping
(even when both players share the same GPU).
However, Capsule does not offer performance isolation,
which is required when players
on the same GPU contend for the same resource.
For example, high GPU consumption by
one player---such as staring at a render-intensive scene
for an extended period---can create contention,
lowering the FPS for other players on that GPU.
Capsule also introduces player fate-sharing by
colocating multiple players on the same GPU.
Thus, if one GPU fails, multiple players suffer.
Our future work can leverage
existing isolation and fault tolerance techniques
to alleviate these limitations.
\autoref{appx:limit} further discusses these techniques.

Capsule's sharing benefits can be further improved.
Capsule's cross-player sharing is inspired
by cross-VM (Virtual Machine) compute and memory sharing,
present since the early days of the cloud
(e.g., by Waldspurger~\cite{waldspurger}
to share memory at the OS page level).
Difference Engine~\cite{diff-engine} further increases
the sharing degree via sub-page-level sharing.
EndRE~\cite{EndRE} applies network-level redundancy-elimination
techniques to reduce bandwidth consumption between cloud endpoints.
Capsule is complementary to these lines of work.
In fact, similar optimizations already exist in our cloud software stack.
At the same time, Capsule can be further improved by applying
these techniques at the game-engine level
(e.g., sub-asset-level sharing, as in to Difference Engine,
and redundancy-elimination in cross-player network streams,
as in EndRE).
Put differently,
Capsule intends to bring already-successful optimizations
in other parts of cloud systems
(e.g., OSes, hypervisors, and storage systems)
to the game-engine level.

\section{Conclusion}

We proposed Capsule:
an efficient in-game-engine player isolation mechanism.
It satisfies player isolation requirements in the cloud.
Our implementation in O3DE
shows that Capsule is application-agnostic.
We ported four existing applications to the Capsule-based O3DE
without application changes.
Our experiments with diverse applications and
diverse hardware specifications
show that Capsule can increase
datacenter resource utilization by
accommodating up to 2.25x more players.
The Capsule design can be adopted by other game engines
to increase cloud datacenter utilization.

\section{Acknowledgments}

We thank all reviewers \&
Xiaofeng Zhang, Wenxiao Zhang, Steven Yuan, Yin Wei,
and Huawei Cloud members for their support.
\appendix
\section{Player State Tracking}
\label{appx:implementation}

\begin{figure}
  \centering
  \includegraphics[width=0.99\linewidth]{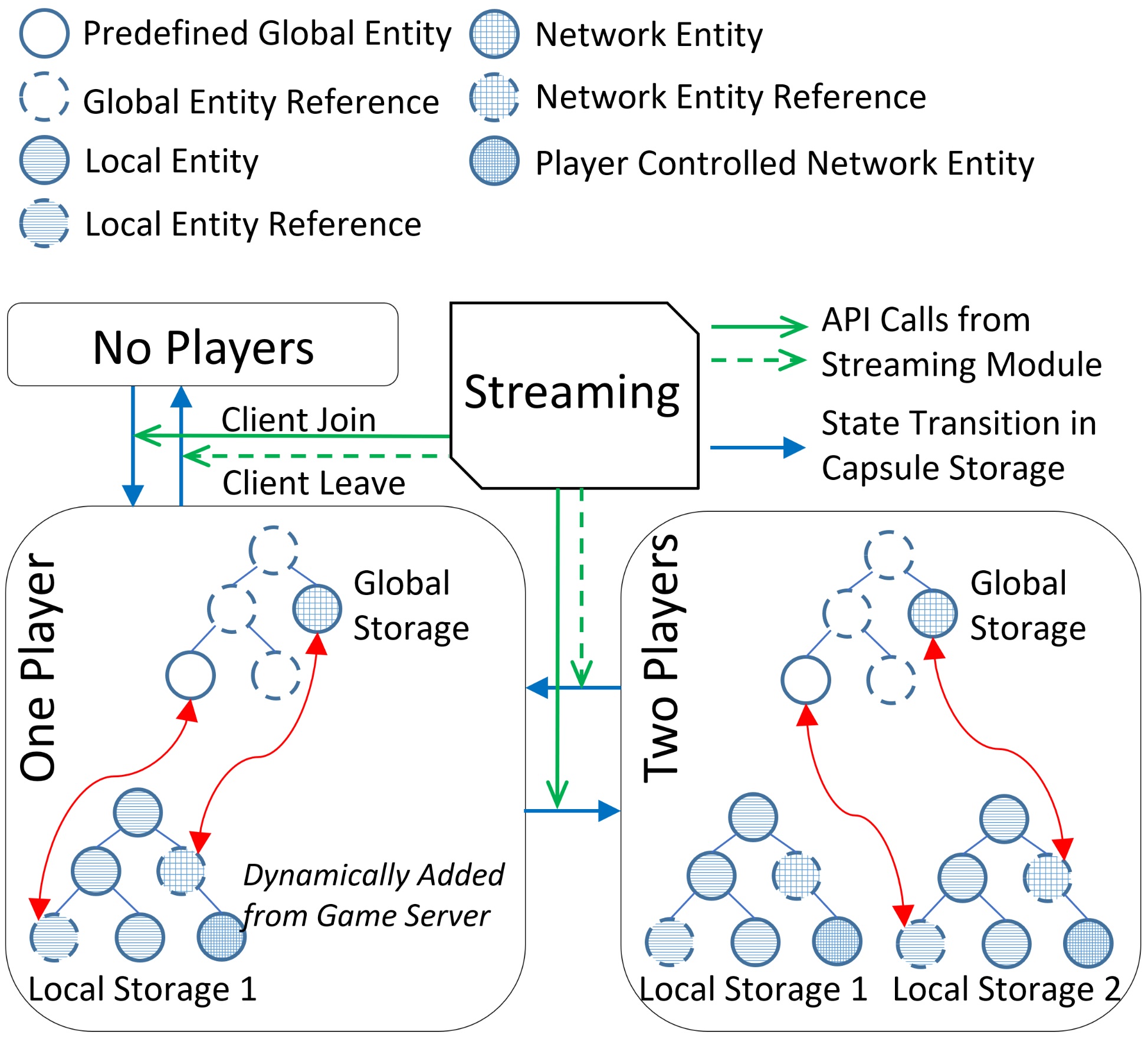}
  \caption{Player entity tracking
  using global and local Capsule storage.
  Global storage is instantiated only once.
  Local storage is created as players join and is destroyed
  when they leave the game.}
  \label{fig:storage}
\end{figure}

\autoref{fig:storage} shows how Capsule handles per-player entity
tracking using global and local storage.
There are seven kinds of constructs, starting with Predefined Global Entity and
ending with Player Controlled Network Entity.
Not all of these constructs are exposed to the game developer.
In fact, game developers design games as they would for
vanilla (non-Capsule) O3DE,
e.g., declare an entity as Network Entity
(such as a museum piece in \hdc)
or Player Controlled Network Entity (such as a player avatar).
Capsule manages entity division into local and global,
as well as reference tracking.
For example, when there are no players,
there are no constructs.
Optionally, Capsule could automatically create and maintain
a game level cache to accelerate game loading when players join.
This cache is invisible to the game.

As~\autoref{fig:storage} shows, players join through
API calls from the Streaming module (see~\autoref{fig:capsule}).
When the API call is made for the first player,
predefined global entities are initialized,
unless they have already been initialized via the cache.
During player join, all entities that the game developer specified
in the initial level are spawned.
For subsequent players,
predefined global entities are not re-instantiated;
instead, references to the original entities are created within
each subsequent player's local storage.
These references ensure that player-local gameplay logic remains consistent
with the vanilla game-level design (for a non-Capsule engine).

When the second player joins,
local entities are duplicated for each player.
A corresponding reference for each duplicate is maintained in the global storage.
This reference mechanism ensures that player-specific gameplay logic 
is consistent with the vanilla design.
Network entities, i.e., entities whose state and logic
must be synchronized through the game server~\cite{mmo-server},
are typically treated as global entities
since they are inherently shared among players.
Such treatment is consistent with legacy multiplayer architecture.

Event tracking follows a flow analogous to entity tracking.
There are global events, such as time-of-day changes in \poh,
and local events, such as bullet inventory decreases
when a player fires in \sample.
These distinctions also apply to the event storage,
which also separates global and local scopes.
Global events may trigger both global and local events,
whereas local events are restricted to interactions
within the same local event storage.
For example, a time-of-day change event affects all players,
whereas a local inventory-update event should remain confined to
that player's local storage.

A notable exception arises with player-controlled network entities,
e.g., the entity that is labeled ``Dynamically Added from Game Server''
in~\autoref{fig:storage}.
This entity is globally visible across all players,
thereby functioning as a global entity,
but its event propagation is restricted to local event storage.
For instance, when a player presses the ``attack'' command, the game generates
animations and state changes for that player without broadcasting to other players.
However, when this action results in interactions with other players
(e.g., a collision), the corresponding events are escalated to the game server
and synchronized globally, making these events visible to other players.
This local-propagation-until-interaction approach ensures
that player inputs remain isolated,
while cross-player interactions are broadcast
across the entire shared game world.

We designed Capsule Storage (\autoref{fig:storage})
and the Capsule architecture
with generality in mind.
Most Capsule changes lie in either the ECS layer
or the input and output systems of the game engine.
This design maintains compatibility with other subsystems,
such as the rendering and physics systems.
Therefore, it is possible---and, in fact, encouraged---
to replace these systems with multi-player-aware alternatives,
without modifying Capsule.
For example, related work on On Surface Caches (OSC)~\cite{osc}
caches and reuses the computed color values
of the same world position across multiple players.
The default O3DE rendering subsystem can be replaced
with an OSC-like alternative to further improve cross-player sharing,
without changing Capsule.

\section{Evaluation Methodology}
\label{appx:method}

In each experiment, the gameplay of each player is \textit{unique},
and is \textit{deterministic} across different hardware.
It is unique because the player walking trajectories
differ within the same experiment.
We record a stochastic trajectory for each player in a separate experiment
and replay that trajectory in Capsule vs. Baseline evaluations.
For example, if there are 4 players in the experiment,
each player trajectory is stochastic.
Stochasticity makes our evaluations unbiased,
i.e., free from the selective benchmarking crime~\cite{bench-crimes}.
The view the player gets in a frame influences the rendering load for that frame,
which in turn influences how much computation
can be shared across multiple players in that frame.
In the biased case, we would have all players have the same view,
or largely overlapping view,
which would unfairly make Capsule outshine the Baseline
because the amount of cross-player sharing is maximized.
By adopting stochasticity, our evaluations are free from such bias.
In fact, stochastic trajectories might undersell Capsule's benefits
because players might inadvertently have less view overlap, hence less sharing,
than they would in realistic production deployments.
We would rather undersell than be biased.
Thus, the Capsule benefits we report in our experiments are conservative.

The determinism across different hardware allows our experiments to stay
true to our claim: hardware differs; everything else stays the same.
When we evaluate different hardware, we replay the prerecorded trajectories
for the first player, the second player, and so on.
Thus, the player generated rendering load is identical across all hardware
configurations while the number of players accommodated might increase
(or decrease) depending on the compute capacity
of the hardware under evaluation.
This exactly is the purpose of the hardware diversity experiments:
evaluate if the Capsule benefits are consistent
across different datacenter hardware.

In general, we chose the strongest viable baseline
and compared it to Capsule.
We evaluated over a wide range of applications and
diverse datacenter hardware
while faithfully replicating our production environment
and remaining bias-free.
Capsule has been implemented for Linux,
but is not as thoroughly performance-evaluated as on Windows.
We believe the results we report here
apply to Linux environments as well.

\section{Evaluation on Diverse Hardware}
\label{appx:eval}

\begin{figure*}
  \centering
  \includegraphics[width=0.99\linewidth]{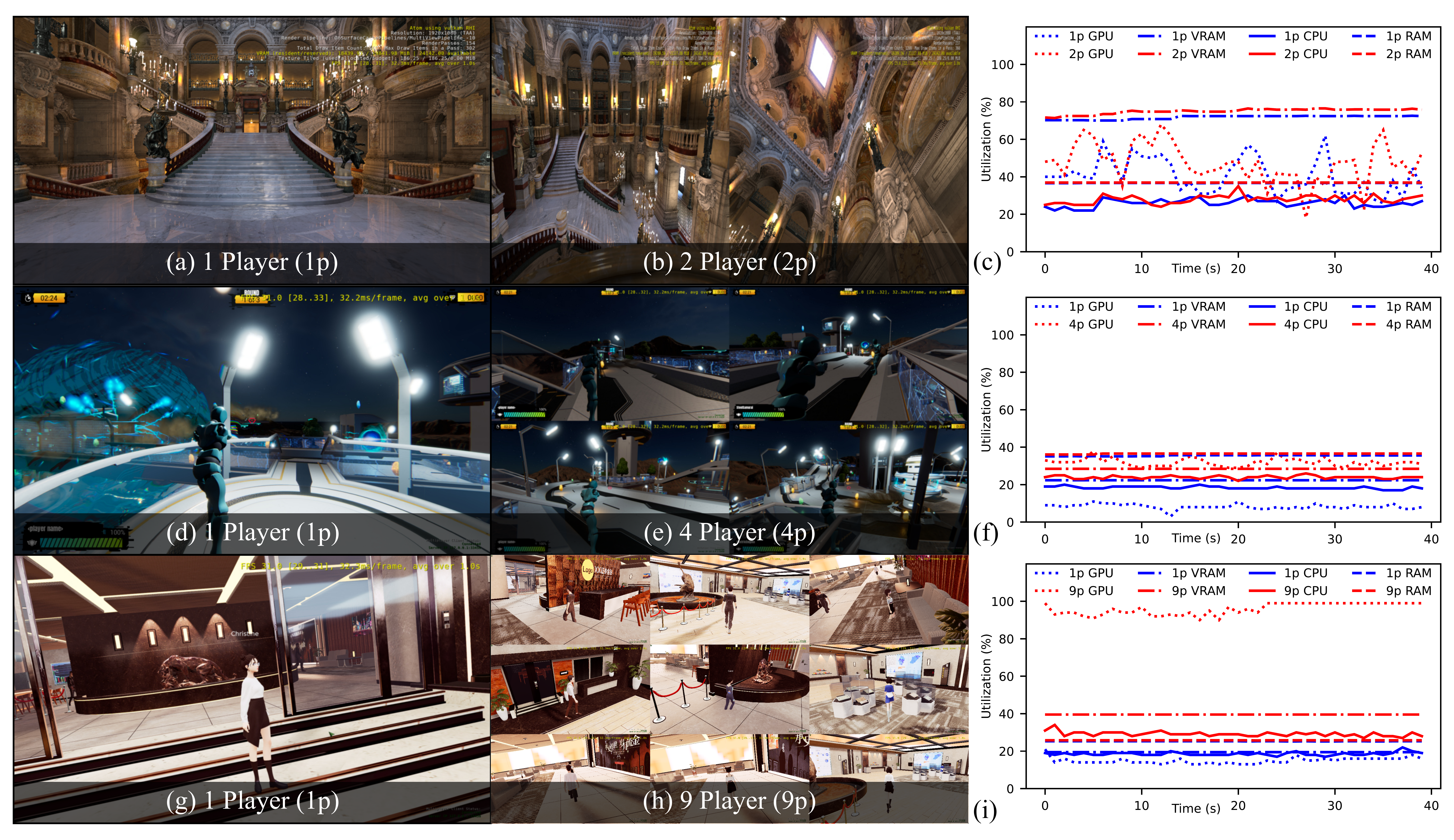}
  \caption{Three examples demonstrate the applicability
  of Capsule to a wide range of graphics-intensive applications:
  1) \textit{high-graphics}: \poh, a digital twin of the Paris Opera House (top);
  2) \textit{medium-graphics}: \sample, a shooting game (middle);
  3) \textit{low-graphics}: \hdc, a digital twin of a museum (bottom).
  Capsule achieves high datacenter resource utilization---GPU, VRAM, CPU, and RAM---while providing
  lightweight, efficient player isolation.}
  \label{fig:teaser-extended}
\end{figure*}

\autoref{fig:teaser-extended} shows results for three applications
on a \singlegpu workstation with the strongest GPU.
\autoref{fig:teaser-extended}(a) and \autoref{fig:teaser-extended}(b) show
the game server's view of \poh with one player (1p) and two players (2p), respectively.
\autoref{fig:teaser-extended}(c) shows resource utilizations for these two environments.
Unlike other applications used in our experiments,
GPU usage in \poh has high variance.
For example, in some small intervals, 2p utilization line is below 1p utilization.
This is partially due to the utilization capturing noise,
but is mostly due to view angle of the players.
For example, between the 25th and 28th seconds,
if two players in the 2p environment happen to view the corner of a wall,
while the single player in the 1p environment views a distant area
with many polygons, 1p GPU utilization will be higher.
However, if we summarize GPU utilization
across the entire 40-second gameplay, 
without focusing on short time intervals,
the average 2p GPU utilization
($\approx$50\%) is $\approx$10\% higher than 1p ($\approx$40\%).
This is expected: with Capsule, the second player imposes a sublinear
$\approx$10\% GPU cost thanks to cross-player sharing.

This delta is smaller, but is more \textbf{significant}
in other resources (VRAM, CPU, RAM).
If the delta is the smallest, i.e., is zero, 1p and 2p lines overlap.
This means that 2p utilization of that resource is identical to that of 1p,
meaning the second player came for free (for that resource).
In~\autoref{fig:teaser-extended}(c),
this is almost the case for VRAM, CPU, and RAM:
the second player has a small extra cost.
Thus, the cost of additional players is sublinear
for these resources.

Capsule is unable to accommodate more than two players in the \poh application.
The FPS drops below the threshold (30) with the third player
due to a CPU bottleneck.
This is not visible in~\autoref{fig:teaser-extended}(c),
i.e., CPU utilization is only $\approx$30\% for 1p as well as 2p.
This is misleading because CPU utilizations are reported and are plotted
for all 8 cores (in \singlegpu workstation, see~\autoref{tab:workstations})
while there is only one main thread for all players,
which runs on a single CPU core.
That thread is the bottleneck.
We can improve Capsule performance by spreading players across CPU cores.
We have not done so, yet, because in most games, GPU is the bottleneck
(as evident in our third application, \hdc).
However, extending Capsule to use different CPU cores
for different players is the right future work
for widening Capsule's applicability to diverse applications.

The sublinear cost conclusion holds in \sample~\cite{o3de-game}
and \hdc applications,
in~\autoref{fig:teaser-extended}(d)--(f) and~\autoref{fig:teaser-extended}(g)--(i),
respectively.
For example, as shown in~\autoref{fig:teaser-extended}(i) for \hdc application,
one player consumes $\approx$18\% of the GPU while nine players consume $\approx$99\%:
a sublinear per-player increase.
Similarly, CPU consumption increases sublinearly: by only $\approx$10 percentage points
with nine players ($\approx$29\%) versus one player ($\approx$19\%).
Note that \sample also suffers from the aforementioned single-thread bottleneck.
\hdc does not, and therefore can achieve over 99\% GPU utilization
for nine players.
These results demonstrate that Capsule, as implemented now,
brings greater benefit to graphics-heavy applications,
i.e., when the GPU is the bottleneck.
Our future work will alleviate the CPU bottleneck.

\begin{figure*}[t]
  \centering
  \includegraphics[width=0.99\linewidth]{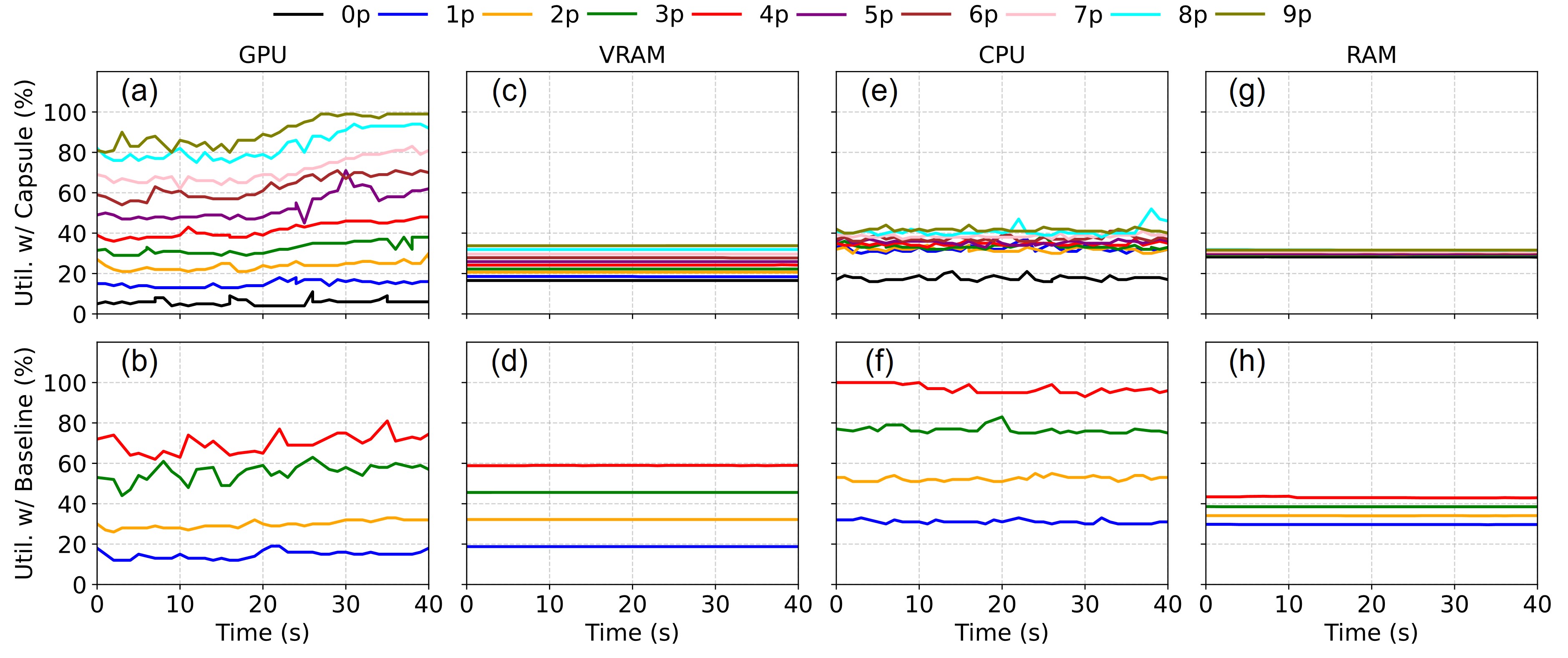}
  \caption{System-wide utilizations with Capsule and Baseline on
  \singlegpu workstation.}
  \label{fig:util1gpu}
\end{figure*}

\begin{figure*}[t]
  \centering
  \includegraphics[width=0.99\linewidth]{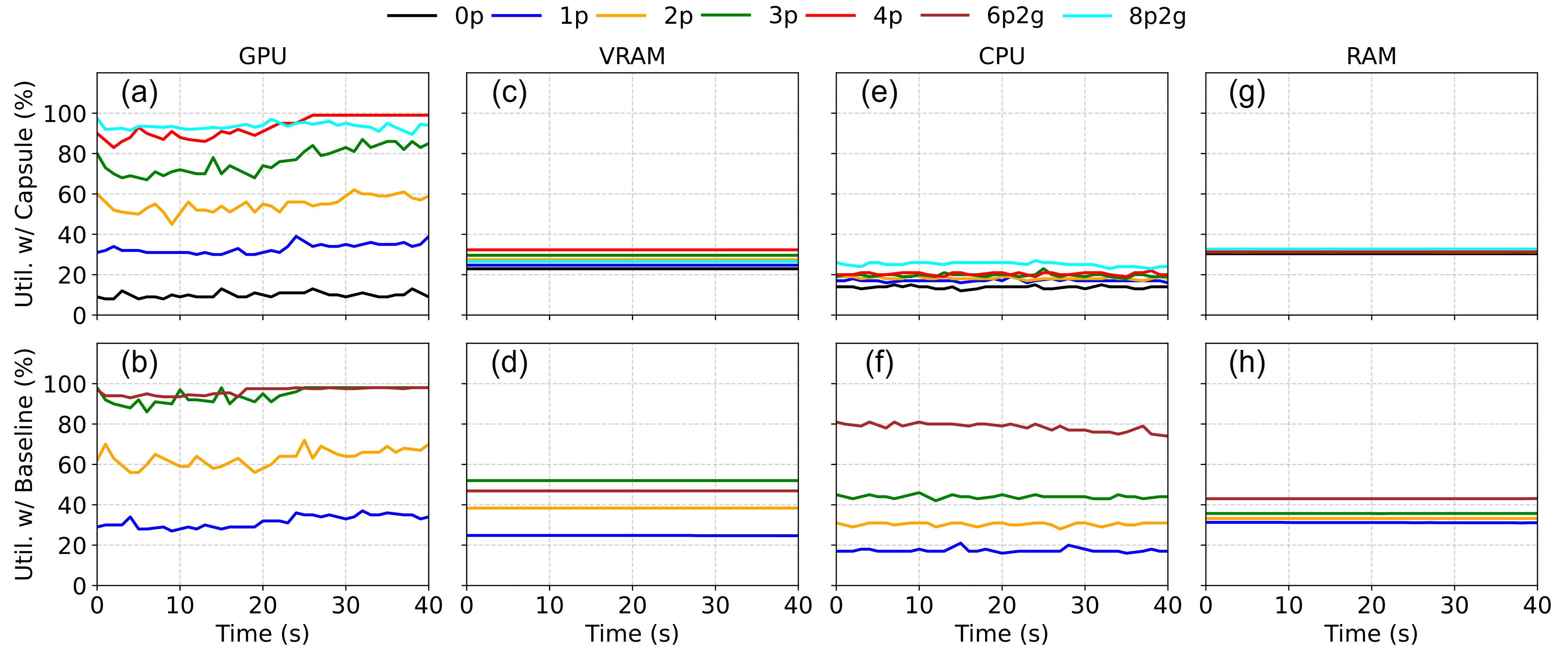}
  \caption{System-wide utilizations with Capsule and Baseline on
  \dualgpu workstation.}
  \label{fig:util2gpu}
\end{figure*}

\autoref{fig:util1gpu} shows a finer view of the experiment in~\autoref{fig:teaser-extended}
by plotting per-second resource utilization during 40-second gameplay
for different numbers of players.
0P represents the initial state where the game server
and the game client are launched, but no players are created yet.
Results from this diversity evaluation are consistent with our earlier
per-player resource footprint experiment,
i.e., Capsule accommodates up to 9 players
while Baseline hits a CPU bottleneck after 4 players.
Thus, 2.25x more players with Capsule.

These 2.25x savings are thanks to Capsule's ability to
multiplex server resources across multiple players.
\autoref{fig:util1gpu}(a)-(b) show GPU utilizations for different numbers of players.
0p line in~\autoref{fig:util1gpu}(a) shows $\approx$10\% GPU utilization when
there are zero players, i.e., only the game server, an empty game client,
and OS background processes are running.
With 1 player (1p),
GPU utilization in Capsule (\autoref{fig:util1gpu}(a)) is similar
to that of Baseline (\autoref{fig:util1gpu}(b)): both are $\approx$20\%.
However, as~\autoref{fig:util1gpu}(b) shows, with the Baseline,
the GPU utilization reaches
up to 37\% with two players,
up to 62\% with three players, and
up to 81\% with four players.
Thus, each player imposes linear, $\approx$20\% GPU overhead.
On the other hand,
as~\autoref{fig:util1gpu}(a) shows, with Capsule,
the overhead is sublinear:
up to 30\% with two players,
up to 40\% with three players,
up to 50\% with four players, and so on until
up to 100\% with nine players.

The reason we stop at nine players (9p) in~\autoref{fig:util1gpu}(a)
is that the GPU becomes the bottleneck after nine players,
which causes the game's FPS to fall below the acceptable threshold (30 FPS).
However, the reason Baseline stops after four players (4p)
is a CPU bottleneck,
not the GPU.
As~\autoref{fig:util1gpu}(f) shows,
the CPU becomes a bottleneck with the fourth player, with the game dropping below 30 FPS.
On the other hand, as~\autoref{fig:util1gpu}(e) shows,
Capsule is able to multiplex CPU resources across multiple players,
even better than it multiplexes GPU resources.
This is because Capsule achieves higher cross-player sharing
for CPU computation than for GPU computation; e.g., Capsule consumes
$\approx$20\% CPU with zero players,
$\approx$32\% with one player,
$\approx$33\% with two players,
and so forth until only $\approx$40\%
(up to 50\%, briefly) with nine players (9p).

A similar trend holds for VRAM (\autoref{fig:util1gpu}(c)--(d))
and RAM (\autoref{fig:util1gpu}(g)--(h)).
\autoref{fig:util1gpu}(c) shows a sublinear VRAM footprint with Capsule
while~\autoref{fig:util1gpu}(d) shows linear VRAM footprint with Baseline.
\autoref{fig:util1gpu}(g) shows a sublinear RAM footprint,
even more sublinear than VRAM,
while Baseline imposes a linear per-player RAM footprint.

We also evaluated Capsule and Baseline with two GPUs on \dualgpu workstation.
For Baseline, we created three processes that use GPU1
and another three processes that use GPU2.
For Capsule, we use a single process,
but in the game engine that runs in that process
four players get assigned to GPU1 and the other four get assigned to GPU2.
Thus, there are eight players in \dualgpu.

\begin{figure*}[t]
  \centering
  \includegraphics[width=0.99\linewidth]{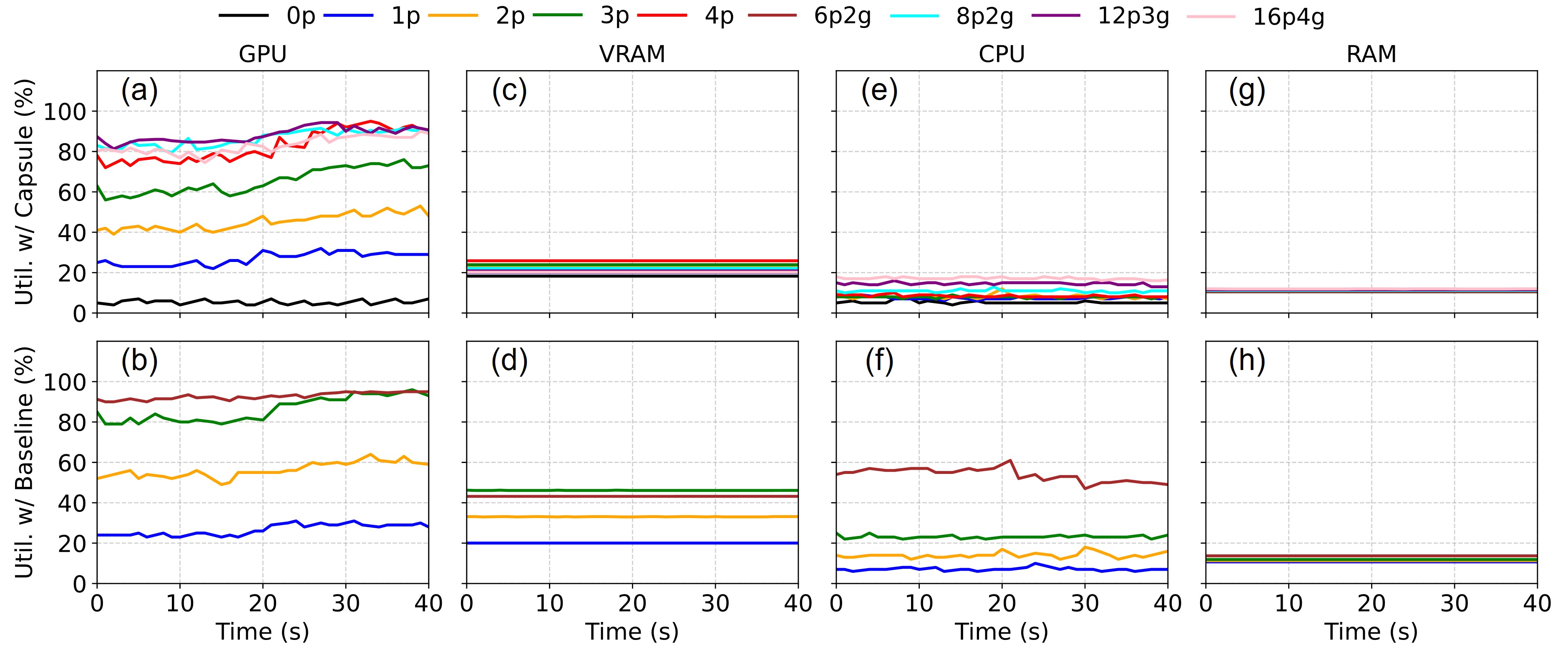}
  \caption{System-wide utilizations with Capsule and Baseline on
  \quadgpu workstation.}
  \label{fig:util4gpu}
\end{figure*}

\autoref{fig:util2gpu} shows the results on \dualgpu workstation
where Baseline hits the GPU bottleneck (unlike in~\autoref{fig:util1gpu}).
However, this time, Capsule brings only 33\% benefit, i.e.,
it supports up to four players
while Baseline supports at most three (with 30 FPS).
Capsule is more effective on \singlegpu workstation
than on \dualgpu workstation because
the former's GPU is significantly (>60\%) stronger:
RTX 4090 with 16,384 CUDA cores of 2.2 GHz clock frequency on the former vs.
RTX 3090 with 10,496 CUDA cores of 1.4 GHz clock frequency on the latter.
Capsule, primarily being a GPU multiplexing technique, has less room to shine
with the weaker GPU.
Thus, the GPU becomes the bottleneck with four players in Capsule
(3 players in Baseline)
on the \dualgpu workstation, whereas that bottleneck is hit with
nine players in Capsule
(4 players in Baseline) on the \singlegpu workstation.

\autoref{fig:util2gpu}(a) shows GPU utilizations with up to
eight players on two GPUs (8p2g) with Capsule and 
\autoref{fig:util2gpu}(b) shows up to six players (6p2g) with Baseline.
Note that GPU and VRAM utilizations in~\autoref{fig:util2gpu}
are the average of both GPUs at any given time.
We capture system-wide utilizations, which include background OS processes.
Thus, in these experiments, GPU1 has higher GPU and VRAM usage compared
to GPU2 due to background processes running on GPU1.
Therefore, when we average across two GPUs, per-player averages
become lower than in the single-GPU experiments.
For example, in~\autoref{fig:util2gpu}(d),
3p VRAM utilization with Baseline is $\approx$53\%
while it is $\approx$48\% in 6p (from the average of two GPUs).
This means 53-48$\approx$5\% VRAM utilization was due to background processes.
Similar background overhead applies to all other multi-GPU experiments.

The findings in~\autoref{fig:util2gpu} are consistent with those
from~\autoref{fig:util1gpu}.
\autoref{fig:util2gpu}(a) shows a sublinear multiplayer resource footprint
for the GPU, i.e., it starts with
$\approx$37\% utilization with one player (1p),
reaching $\approx$100\% on a single GPU with four players (4p),
and reaching $\approx$100\% on two GPUs with eight players (8p2g).
On the other hand, with Baseline (\autoref{fig:util2gpu}(b)),
GPU utilization is
$\approx$37\% with one player (1p),
reaching $\approx$100\% with three players on a single GPU (3p),
and reaching $\approx$100\% with six players on two GPUs (6p2g).
This sublinearity is even more evident for VRAM
(\autoref{fig:util2gpu}(c)--(d)),
i.e., utilization grows by only $\approx$9 percentage points with Capsule
($\approx$22\% in 1p vs. $\approx$31\% in 4p in~\autoref{fig:util2gpu}(c))
versus $\approx$28 percentage points with Baseline
($\approx$25\% in 1p vs. $\approx$53\% in 3p in~\autoref{fig:util2gpu}(d)).

As~\autoref{fig:util2gpu}(e)--(h) show,
additional players also have
sublinear CPU and RAM footprints with Capsule, as expected.
For example, Baseline requires an additional \textbf{$\approx$60} percentage points of CPU
($\approx$20\% in 1p vs. $\approx$80\% in 6p2g in~\autoref{fig:util2gpu}(f))
and \textbf{$\approx$10} percentage points of RAM
to go from one player to six players
($\approx$32\% in 1p vs. $\approx$42\% in 6p2g in~\autoref{fig:util2gpu}(h)).
However, Capsule requires less than \textbf{11} additional percentage points of CPU
(vs.\ \textbf{$\approx$60} for Baseline)
($\approx$18\% in 1p vs. $\approx$29\% in 8p2g in~\autoref{fig:util2gpu}(e))
and \textbf{$\approx$4} (vs.\ \textbf{$\approx$10} for Baseline) additional percentage points of RAM 
(32\% in 1p vs. 35\% in 8p2g in~\autoref{fig:util2gpu}(g))
to go from one player to eight players.

\autoref{fig:util4gpu} shows the results on \quadgpu workstation.
The number of players hosted in Baseline and Capsule is similar
to that on the \dualgpu workstation, as the workstations have comparable GPUs,
i.e., RTX 3090 and \fortis yield comparable performance
in the \hdc application.
\quadgpu workstation has four GPUs and has a very powerful CPU.
We allocate players across multiple GPUs.
Each GPU can accommodate up to four players with Capsule
and only three players with Baseline (with 30 FPS).
Thus, as shown in~\autoref{fig:util4gpu}(a), Capsule can host
up to 16 players on four GPUs (16p4g) before the system hits a GPU bottleneck.
 
(Actually, with Baseline,
we should have been able to host up to 12 players on four GPUs,
but we show only up to six players in~\autoref{fig:util4gpu}.
We actually were able to host eight players.
The GPU driver crashes with an unexpected error when launching the ninth player.
We exclude the third, partially utilized GPU from~\autoref{fig:util4gpu}
to avoid inconsistency in averaging.
As explained earlier, we report average GPU utilizations
for our multi-GPU workstations.
This assumes that all GPUs host an equal number of players.
This is not the case when the third GPU hosts only two players
(before GPU crash)
while there are three players on the first GPU and three on the second.
Therefore, in~\autoref{fig:util4gpu},
we exclude (two players on) GPU3 from our calculations
and just average the utilizations for six players (6p) on GPU1 and GPU2.
We believe the conclusions we draw are still valid after GPU3 exclusion
because our conclusions rely on cross-GPU averages.
Even if GPU3 (and GPU4) did not crash,
we would still have three players on those GPUs,
with Baseline accommodating one fewer player per GPU than Capsule,
just as on GPU1 and GPU2.
Thus, our conclusions would still hold.)

As~\autoref{fig:util4gpu}(a)-(b) show,
additional players consume sublinear GPU resource with Capsule.
For example, Baseline requires an additional \textbf{$\approx$65} percentage points of GPU utilization
($\approx$30\% in 1p vs. $\approx$95\% in 6p2g in~\autoref{fig:util4gpu}(b))
to go from one player to six players,
while Capsule requires \textbf{$\approx$67}
(vs.\ \textbf{$\approx$65} for Baseline)
additional percentage points of GPU utilization
($\approx$30\% in 1p vs. $\approx$97\% in 16p4g in~\autoref{fig:util4gpu}(a))
to go from one player to 16 players.
Similar sublinearity holds for VRAM, CPU, and RAM:
\begin{itemize}
  \item \textbf{VRAM}: Baseline requires an additional \textbf{$\approx$26} percentage points of VRAM
($\approx$20\% in 1p vs. $\approx$46\% in 6p2g in~\autoref{fig:util4gpu}(d))
to go from one player to six players,
while Capsule requires \textbf{$\approx$2}
(vs.\ \textbf{$\approx$26} for Baseline)
additional percentage points of VRAM
($\approx$20\% in 1p vs. $\approx$22\% in 16p4g in~\autoref{fig:util4gpu}(c))
to go from one player to 16 players.

  \item \textbf{CPU}: Baseline requires an additional \textbf{$\approx$51} percentage points of CPU utilization
($\approx$8\% in 1p vs. $\approx$59\% in 6p2g in~\autoref{fig:util4gpu}(f))
to go from one player to six players,
while Capsule requires \textbf{$\approx$11}
(vs.\ \textbf{$\approx$51} for Baseline)
additional percentage points of CPU utilization
($\approx$8\% in 1p vs. $\approx$19\% in 16p4g in~\autoref{fig:util4gpu}(e))
to go from one player to 16 players.
  
  \item \textbf{RAM}: Baseline requires an additional \textbf{$\approx$6} percentage points of RAM
($\approx$10\% in 1p vs. $\approx$16\% in 6p2g in~\autoref{fig:util4gpu}(h))
to go from one player to six players,
while Capsule requires \textbf{$\approx$2}
(vs.\ \textbf{$\approx$6} for Baseline)
additional percentage points of RAM
($\approx$10\% in 1p vs. $\approx$12\% in 16p4g in~\autoref{fig:util4gpu}(g))
to go from one to 16 players.

\end{itemize}

We also evaluated Capsule in a multi-server setup by connecting the
\dualgpu and \quadgpu workstations in a cluster.
We were able to replicate our results from
\autoref{fig:util2gpu} and~\autoref{fig:util4gpu},
i.e., 8 players on the former workstation and 16 players on the latter one.
All 24 players were connected to the same game server
and they had smooth 30 FPS gameplay, as expected.
Thus, we believe Capsule's sublinearity benefits are
applicable to large-scale datacenters.

\section{Limitations and Future Work}
\label{appx:limit}

The transparency requirement (\textbf{R1}) has two aspects:
\textit{functional} and \textit{performance}.
The functional aspect refers to the gameplay experience,
e.g., if two players are in the same game session and
are colocated on the same engine and the same GPU,
one player jumping should not interfere with the other player's jump.
This is the essential part of the \textbf{R1}.
Capsule and process-level isolation,
which we used as the baseline,
satisfy this requirement.
There is also a performance aspect,
e.g., one player exhaustively using GPU resources through frequent jumps
should not reduce FPS for the second player.
This is often called the noisy-neighbor problem in the cloud~\cite{deepdive}.
Process-level isolation does not satisfy this requirement.
Neither does Capsule, at least as implemented now.

Performance transparency, or performance isolation,
in the current Capsule implementation is achieved in an indirect way:
player rate limiting after worst-case analysis.
When the game is deployed in the cloud,
the game goes through offline performance profiling.
The profiler outputs the FPS range this specific GPU sustained
during an exhaustive gameplay,
similar to code-coverage-based analysis
in software engineering~\cite{coverage}.
We then perform worst-case analysis, i.e.,
derive the maximum number of players this GPU can host in the worst case,
which is when all players perform graphics-heavy operations, concurrently.
For example, if profiler outputs FPS=[130-150] range after the offline analysis,
and the game developer requires at least 30 FPS for smooth gameplay,
we deploy at most 130/30$\approx$4 players on this GPU.
We repeat offline profiling and analysis for each GPU flavor
in the cloud.
Some GPU flavors might get disqualified altogether for
this application if they cannot
offer the minimum developer-required FPS even for a single player.

Although indirectly achieving performance transparency works,
in the future,
we would like to explore ways to achieve it directly.
In this exploration,
our guiding principle would be maintaining lightweightness (\textbf{R3})
because unless done with care,
stronger performance isolation mechanisms
impose non-trivial additional overhead,
as commonly known as the \textit{virtualization tax}~\cite{no-hype}.

Capsule also introduces player fate-sharing,
which might degrade player fault tolerance.
For example, when a server GPU fails, all players running on the game engine
hosted on that GPU also fail.
Process-level isolation also suffers from this limitation.
Fate-sharing is not about a player causing a GPU failure
(or any system component failure),
but is about hardware failures themselves.
Without Capsule, there is only one player per game engine,
one engine per GPU, and the GPU failure impacts only that single player.
However, fate-sharing is an inherent problem,
i.e., the moment we decide to multiplex GPU resources across players
we accept fate-sharing; just as students share a bus ride every day:
a broken bus delays everyone's school arrival.
Fate-sharing is also common and inevitable in the cloud:
VMs, containers, lambdas all suffer it,
e.g., when a server fails, all VMs hosted on that server fail.
The disadvantages of fate-sharing can be ameliorated by other mechanisms,
such as player migration with player state replication,
as was done for VMs over 20 years ago~\cite{migration}.

\bibliographystyle{eg-alpha-doi} 
\bibliography{egbibsample}

\end{document}